\documentclass[pre,floatfix,twocolumn,showpacs,a4paper]{revtex4}

\usepackage{graphicx}
\usepackage{dcolumn}
\usepackage{amsmath}
\usepackage{amssymb}
\usepackage{tabularx}
\usepackage{epsfig}
\usepackage{mathrsfs}
\usepackage{rotating}

\begin{document}
\title{Master-equation analysis of accelerating networks}
\author{David M.D. Smith$^{1,2,3,4}$}
\email{d.smith3@physics.ox.ac.uk}
\author{Jukka-Pekka Onnela$^{2,3,5}$}
\author{Nick S. Jones$^{2,3,4}$}
\affiliation{$^{1}$Centre for Mathematical Biology, Oxford University, Oxford OX1 3LB, U.K.}
\affiliation{$^{2}$Department of Physics, Clarendon Laboratory, Oxford University, Oxford OX1 3PU, U.K.}
\affiliation{$^{3}$CABDyN Complexity Centre, Oxford University, Oxford OX1 1HP, U.K.}
\affiliation{$^{4}$Oxford Centre for Integrative Systems Biology, Department of Biochemistry,
Oxford University, South Parks Road, Oxford, OX1 3QU, U.K.}
\affiliation{$^{5}$DBEC, Helsinki, University of Technology, P.O. Box 9203, FIN-02015 HUT, Finland.}

\date{\today}
\begin{abstract}
In many real-world networks, the rates of node and link addition are time dependent. This observation motivates the definition of
\emph{accelerating networks}. There has been relatively little investigation of accelerating networks and previous efforts at analyzing their degree distributions have employed mean-field techniques. By contrast, we show that it is possible to apply a master-equation approach to such network development. We provide full time-dependent expressions for the evolution of the degree distributions for the canonical situations of random and preferential attachment in networks undergoing constant acceleration. These results are in excellent agreement with results obtained from simulations. We note that a growing, non-equilibrium network undergoing constant acceleration with random attachment is equivalent to a classical random graph, bridging the gap between non-equilibrium and classical equilibrium networks.
\end{abstract}
\maketitle

\section{Introduction}
In many real-world evolving networks the rates of node and link addition differ, a phenomenon that gives rise to the concept of \emph{network acceleration}~\cite{GM, AcceleratingNetworks, Zhang:2007,  
Sen}. Examples include the Internet~\cite{internet, Internet2, Vazquez2}, the World Wide Web~\cite{WWW} and collaboration  
networks~\cite{collaboration} and these have prompted some recent scrutiny~\cite{GM, AcceleratingNetworks, Zhang:2007, Sen}. However, the majority of network research to date has overlooked  
acceleration, with many non-equilibrium network models focusing on the rather specific scenario of adding a fixed (expected)  
number of nodes  at each timestep with a fixed (expected) number of links to the existing network~\cite{NewmanReview, Dorogovtsev,  
ABReview}.

Accelerating networks also appear in biology. The evolution of gene regulatory networks has been interpreted as a form of network acceleration~\cite{GM, mattick:2005}. Studies have shown that the number of regulatory genes scales quadratically with genome size in prokaryotic microorganisms \cite{Nimwegen, Croftetal}. In these networks, the number of links also scales quadratically with the number of nodes (operons/genes)~\cite{mattick:2005}. Indeed, gene regulatory networks have provided significant motivation for the investigation and definition of accelerating networks~\cite{GM, mattick:2005, Croftetal,  Failed, SizeConstraints}. Gagen and Mattick proposed that the apparent accelerating nature of gene regulatory networks in prokaryotic organisms imposes an inherent size constraint on the system and this is consistent with empirical observations~\cite{SizeConstraints, Ahnert}. 

For an evolving (unweighted) network, acceleration is related to the rates of link and node addition~\footnote{The definitions we use here can be easily extended to incorporate weighted links~\cite{AcceleratingNetworks}.}. Consider a network comprising $N(t)$ nodes and $M(t)$ links at some time $t$. We now add $n(t)$ nodes and $m(t)$ links to the system such that, at the next timestep, $N(t+1)=N(t)+n(t)$ and $M(t+1) = M(t)+m(t)$.
Some previous investigations of accelerating networks have assumed that the rate of node addition is $n(t) = 1$ and that the  
expected rate of link addition is given by some power-law function of time $\langle m(t) \rangle \propto  
t^{\beta}$, where the exponent $\beta$ describes the nature of the acceleration~\cite{GM, Zhang:2007, Sen}. Indeed, Gagen and  
Mattick defined network acceleration in terms of this parameter~\footnote{The definition of Gagen and Mattick suggested the  
functional form $\langle m(t) \rangle \propto N(t)^{\beta}$ which, for the addition of one node per timestep, has $N(t)\sim t$ 
and this is the scenario they analyzed in Ref.~\cite{GM}.}. However, Smith \emph{et al.} proposed a more conventional definition of network  
acceleration that is a time-dependent property of the system~\cite{AcceleratingNetworks}.  
For an unweighted network, acceleration is simply related to the rates of node and link addition, more specifically, the time derivative of the velocity-like quantity $m(t)/n(t)$. For a deterministic system (as in an empirically observed network), network acceleration can be written~\cite{AcceleratingNetworks}	
\begin{eqnarray}\label{eqn:acc1}
a(t) &=& \frac{M(t+1)-M(t)}{N(t+1)-N(t)} -\frac{M(t)-M(t-1)}{N(t)-N(t-1)}\nonumber\\
{}&=&\frac{m(t)}{n(t)}-\frac{m(t-1)}{n(t-1)}.
\end{eqnarray}
When considering a stochastic system, the definition is more delicate~\cite{AcceleratingNetworks}. For the scenario in which the  
number of added nodes per timestep is constant ($n(t)\equiv n$) \emph{and} the expected number of links added per timestep is independent  
of the particular number added at the previous timestep, we can rewrite Eq.~(\ref{eqn:acc1}) for a stochastically evolving accelerating network  
as~\cite{AcceleratingNetworks}
\begin{eqnarray}\label{eqn:acc2}
\langle a(t) \rangle &=& \frac{\langle m(t)\rangle - \langle m(t-1)\rangle}{n}.
\end{eqnarray}
This definition of network acceleration was supported by a case study of the evolution of Wikipedia that undergoes different accelerating regimes throughout its evolution~\cite{AcceleratingNetworks} and this is the definition we shall adopt in the following.

In Section~\ref{sec:limitations} we review the use of an existing mean-field technique to analyze accelerating networks and discuss its limitations. In Sections~\ref{sec:constaccRA} and \ref{sec:constaccBA} we consider an alternative master-equation method and demonstrate its relevance by applying it to the scenarios of constant acceleration with random and preferential  
attachment. We  derive full, time-dependent solutions for the evolution of the degree distributions of these networks and compare them
to simulated networks. We conclude in Section \ref{sec:conc}.

\section{Limitations of some mean-field techniques}\label{sec:limitations}
For non-accelerating, non-equilibrium networks there are two conventional methods to obtain degree distributions: the \emph{mean-field} and \emph{master-equation} approaches~\cite{NewmanReview, Dorogovtsev, ABReview, BarabasiDeriv2}. The former assumes that both node degree and time are continuous and the latter uses a continuous time approximation. 

The mean-field approach tracks the evolution of the degree of an individual node throughout the network's evolution. The analysis consists of two distinct stages. The first stage is to derive the expected degree of an individual node in the network at a particular time, based upon the time at which the node was added. At each timestep, a single node is added to the system. The process assumes that a node's degree is continuous \emph{and} that links are added continuously throughout the network's evolution. The expected rate of change of the degree of an individual node is related to the probability of it receiving links throughout the dutation of the network's evolution. At any given stage during the network's evolution, a node added at a particular time will have a certain expected degree. This is used to generate a continuous expected degree as a function of the time at which nodes are added to the system. The inverse of this function provides a time of addition to the system corresponding to that expected degree. The second part of the process is to derive a continuous degree distribution for the resulting network. In many growing network models, old nodes tend to have higher degrees as they have longer to accumulate new links. Consider selecting a node at random in such a network. The probability that it has expected degree greater than some value $k_{\alpha}$ is simply the probability that it was added before the time corresponding to expected degree $k_{\alpha}$. A cumulative degree distribution is thus obtained by assuming the time at which a randomly selected node was added to the network will be uniformly randomly distributed across the timescale of the network's evolution. The continuous degree distribution is then derived by differentiating the cumulative degree distribution.

This was the technique used by Barab\'asi {\it et al.}~\cite{Barabasi,BarabasiDeriv2} and it has also been applied to accelerating networks  
\cite{Zhang:2007, collaboration, topology, tunabledegree}. This method was also adopted by Gagen and Mattick for analyzing accelerating networks~\cite{GM}. However, this technique can be inappropriate in certain situations. To highlight the shortcomings of this approach, we consider  
its application to the scenario of constant acceleration such that the acceleration is $a(t) \equiv a$ ~\footnote{In the investigation of Gagen and Mattick, this scenario was described as a  `hyper-accelerating' network~\cite{GM}.} and that the continuous (deterministic) rate of link addition is $m(t) = at$. For an initial seed comprising $N(0)$ nodes and $M(0)$ links, the total number of nodes in the system is $N(t)=N(0) + t \sim t$, the total number of links is $M(t) = M(0)+\int_{0}^{t} m(t) \textrm{d}t \sim \frac{at^2}{2}$ and the mean degree of the network is $2 M(t) /N(t) \sim at$ (where $\sim$ means asymptotically equal to). Consider the case of random attachment in which all new (undirected) links are made between newly introduced nodes and randomly selected nodes within the existing network and no connections are established between existing nodes (leading to the constraint $a(t)< 1$ as the number of newly introduced links cannot exceed the number of existing nodes). Using the mean-field technique, the continuous rate of change of the (continuous) expected degree of some node $\alpha$ can be expressed in terms of the probability that it acquires new links~\cite{BarabasiDeriv2}:
\begin{eqnarray}\label{eqn:meanfieldrandom1}
\frac{\textrm{d}\langle k_{\alpha}(t)\rangle}{\textrm{d}t} &\sim& \frac{ m(t)}{t}.
\end{eqnarray}
Integrating from the time at which the node was added to the system $t_{\alpha}$ to time $t$ yields the expected degree of  
this node:
\begin{eqnarray}\label{eqn:meanfieldrandom2}
\langle k_{\alpha}(t) \rangle &\sim& at.
\end{eqnarray}
The expected degree of node $\alpha$ is asymptotically independent of the time $t_{\alpha}$ at which it was added so the expected degree of all nodes is $\langle k(t) \rangle  \sim at$. It is argued in Ref.~\cite{GM} that the continuous degree distribution derived from the mean-field approach for this system is, therefore, a  delta function at degree $at$.

For the scenario of preferential attachment, the ends of the new links associated with new nodes connect to existing nodes preferentially according to their degree. Again, using the mean field technique, the  
rate equation describing the evolution of the degree of some node $\alpha$ at time $t$ for this process can be expressed as 
\begin{eqnarray}\label{eqn:meanfieldpref1}
\frac{\textrm{d}\langle k_{\alpha} (t) \rangle}{\textrm{d}t} &\sim& \frac{m(t) k_{\alpha}(t)}{2M(t)}.
\end{eqnarray}
Integrating between the appropriate limits provides the same expected degree $\langle k_{\alpha}(t) \rangle \sim at$ which is  
asymptotically independent of the time the node was added and, consequently, the same for all nodes in the system. Again, it is argued in Ref.~\cite{GM} that the mean-field technique produces a delta function for the continuous degree distribution at $at$.  

Gagen and Mattick provide a method to retrieve a discrete degree distribution from a continuous one~\cite{GM}. However,  
because the mean-field technique provides identical continuous degree distributions, the outcome will be the same for both random  
and preferential attachment with constant acceleration (equations (18) and (31) in \cite{GM}). In practice, however, simulations  
of these networks generate very different degree distributions as illustrated in Fig.~\ref{fig:prefrandattach}. Concerns over the above treatments  motivate the approach below.

\section{Random attachment with constant acceleration}\label{sec:constaccRA}
We now employ a master-equation method to derive the degree distribution of a random attachment network undergoing constant acceleration.  
The method was introduced by Krapivsky \emph{et al.}~\cite{KrapivskyDeriv} and independently by Dorogovtsev {\it et al.}~\cite{DorogovtsevDeriv} to derive the steady-state degree distribution of the Barab\'asi-Albert preferential attachment mechanism~\cite{NewmanReview}. A similar technique based upon distribution kinetics has been applied to accelerating networks~\cite{Jeon} and, recently, a master-equation method was developed to investigate the emergence of correlations within evolving networks~\cite{LinkSpaceNetworkAnalysis}. Justifications for using such an approach are discussed in Refs.~\cite{history} and ~\cite{Boll}. 

Consider the scenario of constant acceleration, $\langle a(t)\rangle  = a$ (assumed $\ll 1$), for a stochastic accelerating  
network in which the rate of node addition per timestep is constant at $n(t)=n = 1$ and the expected number of added  
(undirected) links per timestep is $\langle m(t) \rangle = at$. Since $N(t)\sim t$ and $a\ll 1$, it follows that the expected number of new links added per pre-existing node per timestep ($\frac{\langle m(t) \rangle}{N(t)}$) is always small.
We consider the example of random attachment and all the newly  
introduced links are established between the new node and the existing network. There are no links formed between existing nodes.  
The process is governed by the attachment probability kernel $\Theta_k(t)$, defined as the probability that a specific, newly  
introduced link connects to an existing node of degree $k$. At some time $t$, there exist $X_k(t)$ nodes of degree $k$ and we wish  
to compute the expected number of nodes of degree $k$ at time $t+1$. The fraction of nodes of degree $k$ is $c_k(t) =  
X_k(t)/N(t)= X_k(t)/[N(0)+t]\sim X_k(t)/t$. It is assumed that the seed component comprising $N(0)$ nodes is small. The master equation for the evolution of $X_k$ can be expressed in terms of the attachment kernel  
and is written
\begin{eqnarray}
\langle X_k  (t+1) \rangle{}& =& X_k (t) + \langle m(t) \rangle\Theta_{k-1}(t)\nonumber\\
{}&{}&-\langle m(t) \rangle\Theta_{k}(t)+P\{m(t)=k\}.
\label{eqn:nodeXmasterACC}
\end{eqnarray}
The second term on the right-hand side reflects the expected number of connections to $k-1$ degree nodes making them nodes of  
degree $k$. The last term is the probability that the new node itself is a node of degree $k$. 

\begin{figure}
\begin{center}
\includegraphics[width=0.45\textwidth]{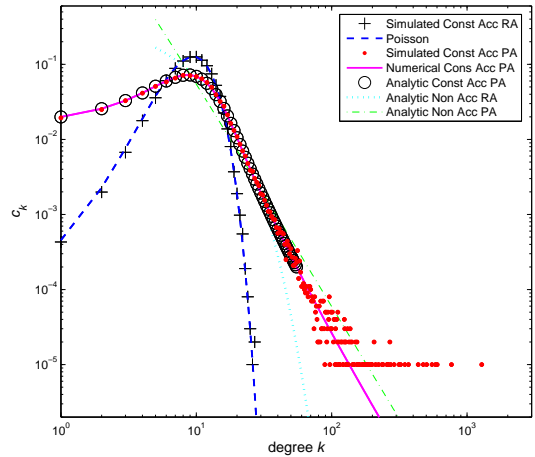}
\caption{ \label{fig:prefrandattach} (Color online) Degree distributions for stochastic random attachment (RA) and preferential attachment (PA) networks undergoing constant acceleration. Each simulation comprises a network grown to $N_{tot}=10^6$ nodes with mean degree of $\lambda  
=10 \approx a N_{tot}$. Also illustrated is the Poisson distribution with the same mean and the analytic expression of  
Eq.~(\ref{eqn:caccfinal}) for the preferential attachment, constantly accelerating network derived in  
Section~\ref{sec:constaccBA}. The numerical implementation simply generates the degree distribution through iterating the master equation,  
Eq.~(\ref{eqn:caccPA1}). The non-accelerating degree distributions for random and preferential attachment (Eqs.~\ref{eqn:ranonacc} and~\ref{eqn:nonaccPA}) with the same mean ($m=5$ links added with each new node) are also illustrated to highlight the effect of accelerating behaviour.}
\end{center}  
\end{figure}

To establish $P\{m(t)=k\}$, it is important that the microscopic mechanism governing the network's evolution be specified. In  
simulating such a network, we proceed as follows: links between the new node and \emph{all} $N(t)$ existing nodes are  
considered and established with probability $a$ satisfying the requirements for the expected number of links for the new node as  
$\langle m(t) \rangle = at$. Interestingly, a network generated via this mechanism will also have the property that every  
possible link between \emph{every pair of nodes} will exist with equal probability $a$. This is equivalent to the classical random  
graph studied by Erd\H{o}s and R\'enyi~\cite{ER} and subsequently Bollob\'as~\cite{Boll2} with identical binomial degree  
distribution. That is, \emph{a growing (non-equilibrium) network undergoing constant acceleration with random attachment is equivalent to the classical random graph,  
bridging the gap between non-equilibrium and classical equilibrium networks}. This is illustrated in Fig.~\ref{fig:prefrandattach}  
where a simulated,  random attachment, stochastic, accelerating network with constant acceleration grown  
to $N_{tot} = 10^6$ nodes is compared to the Poisson degree distribution $c_k = \frac{e^{-\lambda} \lambda^k}{k!}$ of the random  
graph~\cite{ABReview} where $\lambda$ is the mean degree of the network. In Fig.~\ref{fig:prefrandattach}, the mean degree is  
$\lambda=10$. The acceleration parameter is solved for by setting $\lambda  = \frac{2M_{tot}}{N_{tot}}\sim \frac{at^2}{t}\sim aN_{tot}$ so the acceleration is $a = \lambda/N_{tot}$ which subsequently determines $\langle m(t)\rangle$.  

To apply a master-equation approach we start by defining the random attachment probability kernel, $\Theta_k(t) =  
X_k(t)/N(t)=c_k(t)$ for an individual new link attaching to any node of degree $k$ within the existing network.  
For the random attachment mechanism outlined earlier, the number of new links added with the new node will be binomially distributed and, if we  
assume that $t$ is large, we can make a Poisson approximation so that $P\{m(t)=k\} = \frac{e^{-a t}(a t)^k}{k!}$. Non-accelerating networks can have a steady-state degree distribution in the long-time limit and often it is this solution which is investigated from the master equations (for example see Ref.~\cite{LinkSpaceNetworkAnalysis}). However, for accelerating networks, the degree distribution evolves and a  
time-dependent solution must be found. The fraction of nodes of degree $k$ is given by $c_k(t) = X_k(t)/N(t) \sim X_k(t)/t$  
and, using the continuous time approximation $\langle X_k(t+1)\rangle-X_k(t)\approx\frac{\textrm{d}  
\left(X_k(t)\right)}{\textrm{d} t} \sim \frac{\textrm{d} \left(tc_k(t) \right)}{\textrm{d} t}$, we can rewrite Eq.~(\ref{eqn:nodeXmasterACC})  
as
\begin{eqnarray}\label{eqn:acc6}
\frac{\textrm{d} \left( tc_k(t) \right)}{\textrm{d} t}&=& a t c_{k-1}(t) -atc_k(t) + \frac{e^{-a t}(a t)^k}{k!}.
\end{eqnarray}
The time-dependent solution of Eq.~(\ref{eqn:acc6}) gives the degree distribution for this model and is simply 
\begin{eqnarray}\label{eqn:poisson}
c_k(t)&=& \frac{e^{-a t}(a t)^k}{k!},
\end{eqnarray}
as would be expected for a random graph with mean degree of $\lambda \sim at$ in the large size limit~\cite{Boll2}. While this master-equation  
approach does not necessarily imply that a given evolving network will converge upon the solution, in practice this is often the  
case~\cite{Evans}. The Poisson solution of Eq.~(\ref{eqn:poisson}) is compared to a simulated network in Fig.~\ref{fig:prefrandattach}.

The effect of the accelerating nature of the random attachment network can be appreciated by comparison with a conventional, non-accelerating scenario. The master-equation analysis, when applied to a non-accelerating random attachment network with $m(t) \equiv m$  ($m$ integer), gives a steady-state (long-time limit) degree distribution for $k\ge m$ of
\begin{eqnarray}\label{eqn:ranonacc}
c_k = \frac{1}{m+1}\left( \frac{m}{m+1} \right)^{k-m}.
\end{eqnarray}
This is illustrated in Fig.~\ref{fig:prefrandattach} for mean degree of $10$ corrsponding to $m=5$ links added with each new node. 
The mean-field technique applied to the non-accelerating scenario produces the (less accurate) steady-state degree distribution $c_k \propto e^{1-\frac{k}{m}}/m$~\cite{BarabasiDeriv2}. 

\section{Preferential attachment with constant acceleration}\label{sec:constaccBA}
In this section, we apply the master-equation method to the scenario of a constantly accelerating network with linear preferential  
attachment. Again, the network evolves through the addition of one new node per timestep, $n(t)=n = 1$ such that, for a (small) initial seed component of $N(0)$ nodes, the total number of nodes is $N(t)=N(0)+t\sim t$ and links are formed between this node and the existing network. With constant acceleration, $\langle  a(t)\rangle  = a$, the expected number of added links per timestep is $\langle m(t) \rangle = at$. For both simulation and  
analysis, a clear description of the microscopic process is necessary. We assume that the desired resulting network is sparsely connected, i.e. the resultant mean degree is significantly less than the size of the network. As such, $\langle m(t)\rangle \ll N(t)$  
(equivalent to $a\ll 1$) and the attachment process can be modeled as a series of Bernoulli trials. All links between the new node  
and existing nodes in the network are considered. The probability of a link being formed between the new node to some existing  
node $\varsigma$ is set to be ${\langle m(t)\rangle k_{\varsigma}}/{2 M(t)}$. By performing this random trial for all nodes within  
the existing network, both preferential attachment and the expected number of new links added $\langle m(t)\rangle= at$ are  
preserved. Nodes with zero links when introduced to the system will remain degree zero throughout. 

Analysis of this process starts by writing the master equation.  For the purposes of analysis, it is assumed that the total number of links in the network is well approximated by its expected value, $M(t)\approx \langle M(t)\rangle$. Using the preferential attachment kernel, we have $\Theta_k(t) =  
kX_k(t)/2\langle M(t)\rangle$ and the master equation for the evolution of the network can be written
\begin{eqnarray}\label{eqn:caccPA1}
\langle X_k(t+1)\rangle&=&X_k(t)+\frac{(k-1)\langle m(t) \rangle X_{k-1}(t)}{2\langle M(t)\rangle} \nonumber\\ {}&{}&
-\frac{k\langle m(t) \rangle X_{k}(t)}{2\langle M(t)\rangle} + P\{m(t)=k\}.
\end{eqnarray}
We derive the probability $P\{m(t)=k\}$ in a similar manner to the random attachment scenario. Consider the nodes of degree $j$ in the  
existing network at time $t$. There are $X_j(t)$ of them. All $X_j(t)$ possible links are considered between the new node and  
these existing nodes of degree $j$ and each is established with probability ${j \langle m(t)\rangle}/{2\langle M(t)\rangle}$.  
Consequently, the number of new links between the new node and what were $j$-degree nodes will be binomially distributed with mean  
${j \langle m(t)\rangle X_j(t)}/{2\langle M(t)\rangle}$. The probability distribution for the degree of the new node will be a  
convolution of all these binomial distributions corresponding to each value of $j$. However, if we assume that $X_j(t)$ is large  
then we can make a Poisson approximation for each of these binomial distributions. The convolution of two Poisson distributions is a Poisson  
distribution with a mean equal to the sum of the means of the two. As such, the probability distribution for  
the degree of the new node can be approximated by a Poisson distribution. Of course, the mean of this is  
simply $\langle m(t) \rangle$ and, consequently, $P\{m(t)=k\}$ is the same for both preferential and random attachment. Recalling that 
$\langle M(t)\rangle \sim \frac{at^2}{2}$ and that the fraction of nodes of degree $k$ is given by $c_k(t) = X_k(t)/N(t) \sim X_k(t)/t$,  the master equation of Eq.~(\ref{eqn:caccPA1}) for the evolution of a constantly accelerating network with preferential attachment can be rewritten as
\begin{eqnarray}
\frac{\textrm{d} \left( tc_k(t) \right)}{\textrm{d} t}&=& (k-1)c_{k-1}(t) -kc_k(t) + \frac{e^{-a t}(a t)^k}{k!}.
\end{eqnarray}
Multiplying both sides by the integrating factor $t^{k}$, this can be further simplified to
\begin{eqnarray}\label{eqn:caccME}
\frac{\textrm{d} \left( t^{k+1}c_k(t) \right)}{\textrm{d} t}&=& (k-1)t^{k}c_{k-1}(t) + \frac{e^{-a t}(a t^2)^k}{k!}.\nonumber\\
{}&{}&{}
\end{eqnarray}
The recursive nature of Eq.~(\ref{eqn:caccME}) means that the solution can be expressed in terms of repeated definite integrals  
back to the expression for $c_1(t)$ . Each of these integrals equals zero when evaluated at the lower limit allowing the time-dependent degree distribution to be written (see Appendix~\ref{sec:derivation} for details) as
\begin{small}
\begin{eqnarray}\label{eqn:caccfinal}
c_k(t)&=&\frac{1}{t^{k+1}}\sum_{x=1}^{k}\frac{(k-1)!}{x!(x-1)!(k-x)!}\int_0^t (t-y)^{k-x}(ay^2)^x e^{-ay} \textrm{d}y \nonumber\\
{}&=&\sum_{x=1}^k \frac{(k-1)!(2x)!(at)^x}{x!(x-1)!(k+x+1)!} {_1F_1}[1+2x,2+k+x,-at],\nonumber\\
{}&{}&{}
\end{eqnarray}
\end{small}
where $_1F_1[\kappa,\rho,\phi]$ is the confluent hypergeometric function of the first kind, or Kummer's function, defined (for  
integer $\kappa$ and $\rho$) as \cite{Abramowitz}
\begin{small}
\begin{eqnarray}
_1F_1[\kappa,\rho,\phi]&=&\frac{(\rho-1)!}{(\rho-\kappa-1)!(\kappa-1)!}\int_0^1 e^{\phi  
z}z^{\kappa-1}(1-z)^{\rho-\kappa-1}\textrm{d}z.\nonumber\\
{}&{}&{}
\end{eqnarray} 
\end{small}

Comparison of the analytic solution Eq.~(\ref{eqn:caccfinal}) is made to a simulated network in Fig.~\ref{fig:prefrandattach}. The simulated network was seeded with two connected nodes and has acceleration $a = 10^{-5}$. 

To highlight the effects of network acceleration, the master-equation analysis, when applied to a non-accelerating preferential attachment network with $m(t) \equiv m$ ($m$ integer), provides a steady-state (long-time limit) degree distribution for $k\ge m$ of~\cite{Dorogovtsev}
\begin{eqnarray}\label{eqn:nonaccPA}
c_k = \frac{2m(m+1)}{k(k+1)(k+2)}.
\end{eqnarray}
This is illustrated in Fig.~\ref{fig:prefrandattach} for mean degree of $10$ corrsponding to $m=5$ links added with each new node.
The mean-field technique produces a power-law degree distribution, $c_k \propto \frac{m^2}{k^3}$, in the steady state for large $k$ ~\cite{BarabasiDeriv2}.

\section{Conclusions}\label{sec:conc}
We have demonstrated that the master-equation method can be applied to accelerating networks, highlighting the importance of  
specifying the microscopic processes taking place within the evolving network under scrutiny. We have provided full time-dependent  
solutions of the evolving degree distributions for random and preferential attachment with constant  
acceleration.
We note that the classical random graph of Erd\H{o}s and R\'enyi can be modeled as a non-equilibrium evolving network,  
more specifically a constantly accelerating network with random attachment.

\textbf{Acknowledgments:} Smith and Jones are supported by EPSRC and BBSRC and Onnela by a Wolfson College Junior Research Fellowship (Oxford). Smith acknowledges funding from the European Union (MMCOMNET) for part of this research. We thank Felix Reed-Tsochas for comments and suggestions and one anonymous referee for bringing Ref.~\cite{Jeon} to our attention.

\appendix
\section{Analytic solution for constant acceleration with preferential attachment}\label{sec:derivation}
As detailed in Section~\ref{sec:constaccBA}, the master equation for the evolution of the degree distribution of a constantly  
accelerating network with preferential attachment is given by Eq.~(\ref{eqn:caccME}). For clarity, it is useful to write the last term on the right hand side as a function:
\begin{eqnarray}
G_x(t) &=&\frac{e^{-a t}(a t^2)^x}{x!},
\end{eqnarray}
such that the master equation can be rewritten as
\begin{eqnarray}\label{eqn:appcaccME2}
\frac{\textrm{d} \left( t^{k+1}c_k(t) \right)}{\textrm{d} t}&=& (k-1)t^{k}c_{k-1}(t) + G_k(t).
\end{eqnarray}
We can separate the variables and integrate Eq.~(\ref{eqn:appcaccME2}) to give
\begin{small}
\begin{eqnarray}
c_k(t)&=&\frac{1}{t^{k+1}} \Bigg\{ \int_0^t (k-1) {t_1}^{k} c_{k-1}({t_1})\textrm{d}{t_1} + \int_0^{t} G_k({t_1})\textrm{d}{t_1} \Bigg\}.\nonumber
\end{eqnarray}
\end{small}
Making the substitution $k=k'+1$ yields
\begin{small}
\begin{eqnarray}
c_{k'+1}(t)&=&\frac{1}{t^{k'+2}} \Bigg\{ \int_0^t k' {t_1}^{k'+1} c_{k'}({t_1})\textrm{d}{t_1} +  \int_0^t G_{k'+1}({t_1})\textrm{d}{t_1} \Bigg\} .\nonumber
\end{eqnarray}
\end{small}	
We now evaluate this expression explicitly for the first few values of $k' = 0,1,2,3$ to generate the degree distribution for  
the nodes with degrees $1\rightarrow4$:
\begin{tiny}
\begin{eqnarray}
c_1(t)&=&\frac{1}{t^2}\int_0^t G_1(t_1)\textrm{d}t_1, \nonumber\\
c_2(t)&=&\frac{1}{t^3}\Bigg\{ \int_0^t\int_0^{t_1} G_1(t_2)\textrm{d}t_2 \textrm{d}t_1    
 + \int_0^t G_2(t_1)\textrm{d}t_1
\Bigg\}, \nonumber\\
c_3(t)&=&\frac{1}{t^4}\Bigg\{\int_0^t\int_0^{t_1}\int_0^{t_2} 2G_1(t_3)\textrm{d}t_3 \textrm{d}t_2 \textrm{d}t_1  \nonumber\\  
{}&{}& + \int_0^t\int_0^{t_1} 2G_2(t_2)\textrm{d}t_2 \textrm{d}t_1 \nonumber\\
{}&{}& + \int_0^t G_3(t_1)\textrm{d}t_1 \Bigg\},\nonumber\\
c_4(t)&=&\frac{1}{t^5}\Bigg\{ \int_0^t\int_0^{t_1}\int_0^{t_2}\int_0^{t_3} 3\times 2 G_1(t_4)\textrm{d}t_4 \textrm{d}t_3  
\textrm{d}t_2 \textrm{d}t_1\nonumber\\
{}&{}&   + \int_0^t\int_0^{t_1}\int_0^{t_2} 2\times3G_2(t_3)\textrm{d}t_3 \textrm{d}t_2 \textrm{d}t_1   \nonumber\\
{}&{}&   + \int_0^t\int_0^{t_1} 3G_3(t_2)\textrm{d}t_2 \textrm{d}t_1 \nonumber\\
{}&{}&   + \int_0^{t} G_4(t_1)\textrm{d}t_1 \Bigg\}.
\end{eqnarray}
\end{tiny}
We tabulate the coefficients and number of repeated integrals for the contribution of each $G_x(t)$ to each degree distribution  
value $c_k$ in Table~\ref{tab:1} using the notation $(f,g)$ such that $f$ represents the coefficient and $g$ the number of  
repeated definite integrations undergone.
\begin{table*}
\begin{center}
\begin{tabular}{c c c c c c c}
{}&          $G_1(t)$    & $G_2(t)$ &$G_3(t)$ &$G_4(t)$&$G_5(t)$ &$G_6(t)$\\
$c_1(t)$&   $(1,1)$      &  $  -  $&   $ -   $&  $ -  $  &$ -    $&$  - $\\
$c_2(t)$&   $(1,2)$      &  $(1,1)$ &   $-   $&  $-  $  &$ -    $&$  - $\\
$c_3(t)$&   $(1\times2,3)$ & $(2,2)$ &$(1,1)$ &  $- $  &$-$&$ - $\\
$c_4(t)$&   $(1\times2\times3,4)$ &  $(2\times3,3)$ &$(3,2)$ &  $(1,1)  $  &$ - $&$ - $\\
$c_5(t)$&   $(1\times2\times3\times4,5)$ & $(2\times3\times4,4)$ &$(3\times4,3)$ &  $(4,2)  $  &$ (1,1)    $&$  -$\\
$c_6(t)$&   $(1\times2\times3\times4\times5,6)$ & $(2\times3\times4\times5,5)$ &$(3\times4\times5,5)$ &  $(4\times5,3)  $  &$  
(5,2)    $&$ (1,1) $\\
\end{tabular}
\caption{\label{tab:1} The contribution of the functions $G_x(t)$ to the values of the degree distribution $c_k(t)$. The notation  
$(f,g)$ is such that $f$ is the coefficient and $g$ is the number of repeated integrals.}
\end{center}
\end{table*}
By inspection from Table~\ref{tab:1}, we can write the solution of the time-dependent degree distribution as
\begin{small}
\begin{eqnarray}\label{eqn:appfull0}
c_k(t)&=&\frac{1}{t^{k+1}}\sum_{x=1}^k\frac{(k-1)!}{(x-1)!}
\underbrace{\int_0^t \cdots  \int_0^{t_{k-x}}}_{k-x+1}G_x(t_{k-x+1})\textrm{d}t_{k-x+1} \cdots \textrm{d}t_{1}.\nonumber\\
{}&{}&{}
\end{eqnarray}
\end{small}
Each integral, when evaluated at the lower limit, is always zero as $G_x( 0)=0$ for all $x$. The upper limits are all equal in value. This allows further streamlining of the expression for the degree distribution  as the following identity holds for the repeated integral~\cite{RepeatedIntegral}:
\begin{small}
\begin{eqnarray}
\underbrace{\int_0^t \cdots  \int_0^{t_{k-x}}}_{k-x+1}G_x(t_{k-x+1})\textrm{d}t_{k-x+1} \cdots \textrm{d}t_{1} 
&=& \int_0^t \frac{G_x(y) (t-y)^{k-x}}{(k-x)!}\textrm{d}y.\nonumber\\
{}&{}&{}
\end{eqnarray}
\end{small}
Substituting back in the expression for $G_x(t)$, the distribution of Eq.~(\ref{eqn:appfull0}) can now be written as in Eq.~(\ref{eqn:caccfinal}).  

\end{document}